\documentclass[notitlepage,prl,aps]{revtex4-1}
\usepackage{epsfig}

\newcommand{\vA}{{\mbox{\boldmath$A$}}}
\newcommand{\vB}{{\mbox{\boldmath$B$}}}

\newcommand{\vk}{{\mbox{\boldmath$k$}}}

\begin{document}

\title{Role of $d$-wave pairing in $A15$ superconductors.}

\author{S. Mukherjee}
\affiliation{NBIA, Niels Bohr Institute, Blegdamsvej 17, 2100 Copenhagen $\emptyset$, Denmark}
\author{D. F. Agterberg}
\affiliation{Department of Physics, University of Wisconsin-Milwaukee, Milwaukee, WI 53211}
\begin{abstract}
We argue that the recent Raman spectroscopy observation of a sharp mode in $s$-wave superconducting V$_3$Si is due to a competing $d$-wave pairing state.
We present microscopic arguments for the origin of this $d$-wave order.
We further argue that the $d$-wave order explains the observed shrinking of the vortex core structure at anomalously low magnetic fields and the
large anisotropy observed in the upper critical field.
\end{abstract}
\maketitle
Competing pairing channels have played a prominent role in theories
of superfluid $^3$He, cuprate, and pnictide superconductors. In
$p$-wave $^3$He, an excited pair state, stemming from a competing
$f$-wave channel, has been observed through interferometry
in an acoustic cavity \cite{davis}. The observation of this $f$-wave
pairing channel leads to predictions of rich physics near impurities,
surfaces, and vortices. In the cuprates, a competing
$s$-wave or $d_{xy}$ order has been invoked to interpret the
properties of surface Andreev bound states \cite{sauls}. More
recently in the superconducting pnictides, many theories find that a
sign changing $s$-wave state and a competing $d$-wave state have
close transition temperatures\cite{kuroki}, leading to the
prediction of a pair collective mode observable through Raman
spectroscopy and a superconducting state that breaks time reversal
symmetry\cite{lee}.

In this work we argue that $A15$ superconductors provide an ideal
candidate for competing superconducting order.
We show it is natural to expect an attractive $d$-wave pairing channel in addition to the attractive $s$-wave pairing channel in $A15$ materials. Specifically, the $d$-wave pairing belongs to the cubic $E_g$ representation. This representation has two degenerate states with the same symmetry properties as $k_x^2-k_y^2$ and $(2k_z^2-k_x^2-k_y^2)/\sqrt{3}$. The primary experimental motivation for exploring this state is Raman spectroscopy. Raman active superconducting gap modes have been seen in Nb$_3$Sn and in V$_3$Si \cite{hackl}. Modes have been found in three symmetries: $A_{1g}$, $E_g$, and $F_{2g}$. The modes in the $E_g$ channel are significantly sharper and occur at lower energies than those in the $A_{1g}$ and $F_{2g}$ channels. We argue that the $E_g$ modes are due to an attractive $d$-wave pairing channel that arises from the unique structure of cubic $A15$ materials. This $d$-wave order does not play an important role in the ground state thermodynamics but, as we show below, it is important for understanding the superconducting state of A15 materials in magnetic fields.

Many $A15$s are type II  superconductors with transition temperature ($T_c$) on the order of 20 K \cite{math}. Their cubic structure (space group $Pm3n$) contains the formulae unit A$_3$B. The B atoms
form a cubic lattice while the A atoms form
three sets of orthogonal chains that run through the faces of the
cube \cite{geller}. These chains play an important role in understanding these materials. For example, interactions on these chains drive the Martensitic cubic to tetragonal transition that appears in some of these materials \cite{bat1,shirane}. Early theories assumed the electronic states on the chains to be one-dimensional \cite{labb,Gorkov}. However, it was later found that this was an oversimplification \cite{matd,mihai}. Nevertheless, the band structure shows strong anisotropies that originate from electrons residing on the chains \cite{jarlborg}. In accord with this, we assume that the superconducting condensation energy is dominated by  electrons on the three orthogonal chains. This naturally leads to {\it three} order parameters $\Psi=(\psi_{x},\psi_{y},\psi_{z})$ corresponding to superconductivity on each of the three chains (we call this the chain basis). In the following we exploit this chain basis to show how a $d$-wave collective mode arises that can account for the Raman spectroscopy data. We then turn to the development of a microscopic model that provides an estimate for the pairing interactions in the $s$ and $d$-wave channels. Finally, we consider the appearance of the $d$-wave order when magnetic fields are applied.

\noindent {\it Ground state and $d$-wave collective mode}: To address the origin of collective modes, we begin with a Ginzburg Landau (GL) theory that is based solely on symmetry arguments. We use the chain basis to write this theory, in which case the GL free energy density is
\begin{widetext}
\begin{eqnarray}
\label{eqn:chain}
f&=&\alpha \sum_{i}|\psi_i|^2 + \epsilon \sum_{i\neq j}\psi_i\psi_{j}^{*} +
\kappa_1\sum_i|D_i\psi_i|^2 + \kappa_2\sum_{i\neq j}|D_i\psi_j|^2 + \kappa_3\sum_{l,i\neq j}(D_l\psi_i)(D_l\psi_j)^{*}
+ \kappa_4\sum_{l\neq i\neq j}(D_l\psi_i)(D_l\psi_j)^{*}\nonumber\\
&+&\beta_1\sum_i |\psi_i|^4
+\sum_{i\neq j}(\beta_2\psi_i^2{\psi_j^{*}}^2+\beta_3|\psi_i|^2|\psi_j|^2)
+\sum_{l\neq i\neq j}(\beta_4\psi_i\psi_j^{*}|\psi_l|^2
+\beta_5(\psi_i\psi_j{\psi_l^{*}}^2+c.c)\nonumber \\ &+&\beta_6\sum_{l\neq i\neq j}((\psi_i+\psi_j)\psi_l^{*}+c.c)|\psi_l|^2) \label{eqn:chain}
\end{eqnarray}
\end{widetext}
where $D_i=-i\partial_i-2eA_i/c$, ${\bf A}$ is the vector potential, and $c.c.$ represents complex conjugate. Depending upon the coefficients $\epsilon$ and $\beta_i$, Eq.~\ref{eqn:chain} allow for four possible ground states:
$\Psi_s\propto(1,1,1)$, $\Psi_{d}\propto(1,e^{i2\pi/3},e^{i4\pi/3})$, $\Psi_{d,1}\propto(1,-1,0)$, or $\Psi_{d,2}\propto(-1,-1,2)$. The state $\Psi_s$ has $s$-wave symmetry ($A_{1g}$) while the states $\Psi_d,\Psi_{d,1},\Psi_{d,2}$ all are $d$-wave, belonging to the $E_g$ representation. The state $\Psi_s$ will be stable if $\epsilon<0$ while one of the states $\Psi_d,\Psi_{d1}$, or $\Psi_{d2}$ will be stable if $\epsilon>0$.  Physically, it is expected that gap on each chain is maximized to maximize the condensation energy. This implies that the states $\Psi_{d,1}$ and $\Psi_{d,2}$ are unlikely to be realized. Of the remaining two states, the most likely is $\Psi_s$ as the preponderance of evidence suggests an $s$-wave ground state in $A15$ superconductors. In the following it is assumed that $\Psi_s$ is the ground state. However, it should be noted that it is possible that $\Psi_d$ is the ground state of some $A15$ superconductors.

We now turn to possible collective modes of $\Psi_s$. Since our order parameter has six complex degrees of freedom, there can exist six collective modes: the amplitude and phase modes for each of the chain gaps. The amplitude modes are likely to be strongly damped since the mode gaps will typically lie near the chain gap values. Furthermore, the phase mode in which all three chain order parameters oscillate in phase will be pushed to the Plasma frequency. This leaves two phase modes as candidates for collective modes. These modes correspond to relative phase oscillations between the chain order parameters and resemble Leggett modes of multiband superconductors \cite{leggett}. To examine these two phase modes in more detail, we follow an approach developed in Ref.~\cite{kumar} in the context of unconventional superconductors. We represent the order parameter as $\Psi=\psi_0(e^{i\theta_1},e^{i\theta_2},e^{i\theta_3})/\sqrt{3}$.
We use the Josephson relation $\hbar \partial\theta_{\alpha}/\partial t = -\mu_{\alpha}$ together with the quasi-conservation laws
$\hbar\partial \mu_{\alpha}/\partial t=(\partial \mu_{\alpha}/\partial N_{\alpha})(\partial F/\partial \theta_{\alpha})$ where
$\mu_{\alpha}$ is the chemical potential, $F$ is the free energy, and $N_{\alpha}$ the particle number corresponding to quasi-particles on chain $\alpha$. This yields a two-fold degenerate relative phase mode with $d$-wave $E_g$ symmetry that has a frequency given by
\begin{eqnarray}
 \omega_0^2=\frac{2}{3\hbar^2N(0)}\left[3|\epsilon|\psi_0^2-(4\beta_2+\beta_4+3\beta_5+2\beta_6)\psi^4_0\right]
\end{eqnarray}
where $N(0)\equiv \partial N_{\alpha}/\partial \mu_{\alpha}$, is the density of states for each of the chains.

In the limit $\epsilon=\beta_2=\beta_4=\beta_5=\beta_6=0$, this gives $\omega_0=0$. In this limit, the chains are uncoupled and the theory has an accidental $U(1)\times U(1)\times U(1)$ symmetry associated with the independent variations of the phase of each chain. The microscopic theory presented later reveals that $4\beta_2+\beta_4+3\beta_5+2\beta_6<0$, so that $\omega_0$ is well defined. If $\omega_0<2\Delta_0$, where $\Delta_0$ is the chain gap value, these modes will be weakly damped and provide a natural explanation for the Raman spectroscopy results of Ref.~\cite{hackl}.

\begin{figure}
 \epsfxsize=3.5 in \center{\epsfbox{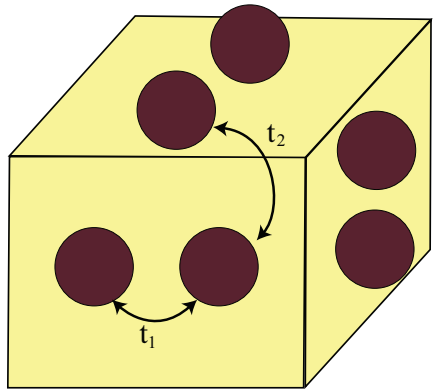}}
\caption{Position of Vanadium atoms in
the unit cell and the corresponding hopping matrix elements included in the Hamiltonian.}
\label{fig:fig1}
\end{figure}

\noindent {\it Microscopic theory}: Now we turn to a microscopic model to examine how an attractive $d$-wave pairing channel can appear in A15 materials and also to determine the phenomenological coefficients in the GL theory.
LDA calculations show that near the Fermi energy, the $3z^2-r^2$ orbitals on the $A$ atoms that form chains along the $z$-direction and symmetry equivalent orbitals for the other two chains directions are the most important \cite{matb}. We therefore use a tight binding Hamiltonian that includes a nearest neighbor (nn)  intrachain hopping ($t_1$) as well as a next nearest neighbor (nnn) interchain hopping ($t_2$) between these orbitals (as shown in Fig.~\ref{fig:fig1}).  This is similar to the approach used earlier by Bhatt \cite{bhatt}. To identify the interactions that lead to superconductivity, we note that there is compelling evidence for strong electron phonon (ep) coupling in $A15$ materials. In particular, phonon anomalies due to strong ep coupling of the chain atom displacements to electrons have been observed \cite{pint} that are in agreement with predictions from LDA calculations \cite{matb}. Additionally, tunneling data on superconducting Nb$_3$Sn reveal features of the phonon spectrum \cite{geerk}, showing that this coupling is responsible for superconductivity. The phonons that couple most strongly to electrons are intra-chain phonons \cite{pint}. We therefore use the following BCS Hamiltonian as a minimal model for $A15$ superconductors
 \begin{eqnarray}
\label{eqn:ham}
\label{eq:ham1}
H=t_1\sum_{\nu\sigma}\sum_{<ij>_{nn}}c_{\nu i\sigma}^{\dag}c_{\nu j\sigma}+t_{2}\sum_{\nu\nu'\sigma}\sum_{<ij>_{nnn}}c_{\nu i\sigma}^{\dag}c_{\nu' j\sigma}-
V_0\sum_{i\nu}c_{i\nu\uparrow}^{\dag}c_{i\nu\downarrow}^{\dag}c_{i\nu\downarrow}c_{i\nu\uparrow},
\end{eqnarray}
where  $c_{\nu i\sigma}$ destroys an electron with spin $\sigma$, with the Wannier function centered on the $i^{th}$ Vanadium atom that transforms as the $\nu$ orbital [$\nu$ corresponds to one of $(3z^2-r^2,3x^2-r^2,3y^2-r^2)$]. The tight binding Hamiltonian is given in more detail in the supplemental notes \cite{sup}. The pairing interaction $V_0$ corresponds to an effective interaction that would arise from ep coupling on the chains alone. The ep coupling gives rise to a local interaction that is retarded in time by an amount $\Delta t\sim 1/\omega_d$, where $\omega_d$ refers to the Debye frequency. This retardation is captured within BCS theory by restricting the energies of the electrons to lie within a cutoff $\omega_d$ of the Fermi energy.

To proceed, we diagonalize the kinetic energy term and this leads to three Fermi surface sheets and a momentum dependent pairing interaction $V_{\alpha,\beta}(\vk,\vk')$ ($\alpha,\beta$ correspond to the different bands). Treating this interaction within a mean-field theory and defining a $\vk$-dependent gap on each of sheets of the Fermi surface, we get the following linearized gap equation

\begin{eqnarray}
\Delta_{\alpha}(\vk)=\ln\left({1.13 \epsilon_D\over k_B T_c}\right)\sum_{\beta}\langle V_{\alpha,\beta}(\vk,\vk')\Delta_{\beta}(\vk')N_{\beta}(\vk')\rangle_{F.S}
\end{eqnarray}

where  $\alpha,\beta$ correspond to the different bands, $\epsilon_{\beta}(\vk')$ is the band energy, $\epsilon_D=\hbar \omega_D$, $N_{\beta}(\vk')=1/|\nabla_{\vk'}[\epsilon_{\beta}(\vk')]|$ is the density of states evaluated at the Fermi energy in the gap equation and the angular bracket implies average over the Fermi surface. To solve this equation we do not assume a $\vk$-dependence for the gap but rather discretise the Fermi surface and explicitly find the eigenfunctions that solve this gap equation. We find that there are only two solutions, an $s$-wave $A_{1g}$ solution and a $d$-wave $E_g$ solution. It is possible to include Coulomb interactions analytically in the limit $t_2/t_1=0$ by using an Anderson-Morel type theory \cite{anderson}. In this limit, we find  that on-site Coulomb interactions have the same effect on both the $s$ and $d$-wave solutions.

\begin{figure}
\epsfxsize=2.5 in \center{\epsfbox{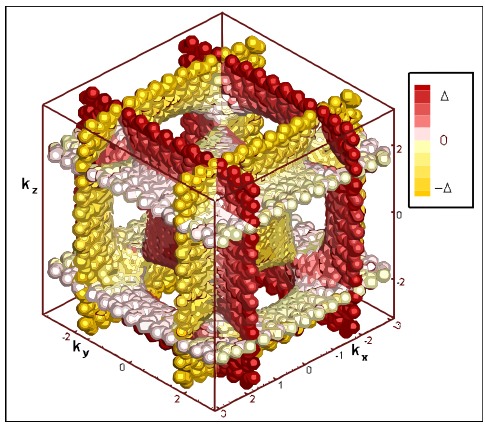}}
\caption{The self-consistent $d_{x^2-y^2}$ like solution for the
$E_g$ symmetry gap function shown on the Fermi surface (for
$t_2/t_1=0.22$). The gap values are represented by the different
colors show in the box on the right. The $(k_x,k_y,k_z)$ are in
units of the inverse lattice spacing.}
\label{fig:fig2}
\end{figure}
In Fig.~2, we show the self-consistent
solution for the $E_g$ gap degree of freedom corresponding to a
$d_{x^2-y^2}$ symmetry. This solution is qualitatively different
than the $k_x^2-k_y^2$ solution expected for a spherical Fermi
surface. In fact, it is more appropriately described as having three
nearly constant gap amplitudes (denoted by $\Delta,0,-\Delta$) which can be interpreted as the three chain gap values.  In
Fig.~3, we show the effective interactions for the $A_{1g}$ and
$E_g$ pairing channels as a function of $t_2/t_1$ while keeping the
Fermi surface volume constant (in the weak-coupling limit
$T_c=\omega_D\exp{\left({-1\over \lambda}\right)}$ where $\lambda$
is the effective interaction found from solving the gap equation).

\begin{figure}
\epsfxsize=2.5 in \center{\epsfbox{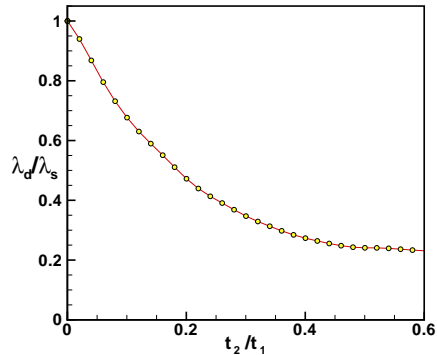}}
\caption{Ratio of eigenvalues $\lambda_d/\lambda_s$ vs $t_2/t_1$.}
\label{fig:fig3}
\end{figure}

\noindent {\it Magnetic Fields}: $A15$ superconductors exhibit anomalous properties in magnetic fields. Notably, these cubic materials have a large anisotropy in the upper critical field ($H_{c2}$) \cite{khlopkin} and they also exhibit
a hexagonal to square vortex lattice transition at anomalously low magnetic fields \cite{Yethiraj}.  This vortex transition is also accompanied by a sharp reduction in the coherence length \cite{sonier,kadano}. Here we examine whether the appearance of $d$-wave order can provide an explanation for these observations.

First we begin with the anisotropy in the upper critical field.  In particular, $H_{c2}$ in V$_3$Si shows an anisotropy that is estimated to be 20 \% at the zero temperature \cite{khlopkin}.
$H_{c2}$ is highest for the field along the $\langle100\rangle$ direction, lower for fields along the $\langle110\rangle$ direction,
and lowest for the field along the $\langle111\rangle$ direction \cite{khlopkin}. As shown in the supplemental notes \cite{sup}, the GL theory of Eq.~\ref{eqn:chain} above predicts that the same anisotropy in $H_{c2}$ as observed experimentally (as described in the previous sentence). Furthermore, this conclusion is {\it independent of the values of the coefficients $\kappa_i$}.

Now we turn to the hexagonal to square vortex lattice transition that occurs at low fields \cite{Yethiraj}. The free energy of Eq.~\ref{eqn:chain} contains too many undetermined coefficients to carry this out. We therefore use the weak-coupling theory described above to find the coefficients in the GL free energy. The details are given in the supplemental notes \cite{sup}.  Of particular relevance, we find that the coefficients $\beta_2, \beta_5$, $k_4$, and $k_3$ are relatively small and we neglect them in the following. The GL theory can then be used to construct a London theory valid at low fields. We take $\psi_i=\psi_0e^{i\phi_i}$, where $|\psi_0|$ takes on the bulk uniform solution value. This yields the following London free energy density
\begin{eqnarray}
f_L&=&f_0 + (\epsilon|\psi_0|^2 + \beta_4|\psi_0|^4)[\cos(\phi_1-\phi_2)+\cos(\phi_2-\phi_3)+\cos(\phi_3-\phi_1)-3]\nonumber\\
&+&\kappa_1|\psi_0|^2\sum_i(\nabla_i\phi_i-2eA_i)^2
+\kappa_2|\psi_0|^2\sum_i(\nabla\phi_i-2e\vA)^2
\end{eqnarray}
where $f_0$ is the free energy in the homogeneous state. Let us define $\phi_1=\theta+\eta_1-\eta_2/\sqrt{3}$, $\phi_2=\theta-\eta_1-\eta_2/\sqrt{3}$, and $\phi_3=\theta+2\eta_2/\sqrt{3}$. We expand $f_L$ to quadratic order in $\eta_1$ and $\eta_2$ (this is valid when $\kappa_1/\epsilon<<a^2$, where $a$ is the spacing between vortices). We then solve for $\theta$, $\eta_1$ and $\eta_2$ to find the following non-local London free energy
\begin{eqnarray}
\label{lon}
f_L&=&{1\over 8\pi}[\vB^2+\lambda_0^2(\nabla\times \vB)^2+{4\kappa_1\over 3|\epsilon|\lambda_0^2}{\kappa_1\over 3\kappa_2+\kappa_1}(\partial_x\partial_yB_z)^2] \label{London}
\end{eqnarray}
where $\lambda_0$ is the penetration depth. Eq.~\ref{lon} has the same form as the non-local London theory used by Kogan \cite{kogan} to examine the hexagonal to square transition. The important point here is that the non-local correction (the third term in Eq.~\ref{lon}) is due to appearance of $d$-wave order and it is possible for this term to become large at relatively low fields, when $\kappa_1/\epsilon \approx a^2\approx B/\Phi_0$ ($\Phi_0$ is the superconducting flux quantum). This would push the hexagonal to square transition to low fields, in agreement with experiment.
However when $\kappa_1/\epsilon > a^2$, the non-local London theory no longer applies. To understand the physics in this regime requires numerical analysis. Some understanding can be found by setting $\epsilon=0$ and $\beta_4|\psi_0|^2=0$. In this case, the three order parameters  $\psi_i$ become decoupled and the vortex cores for each component are smaller than that of the pure $s$-wave theory. This is consistent with the observed shrinking of the vortex core \cite{sonier,kadano}. While more analysis is required, the above shows that the low field hexagonal to square transition and shrinking of the vortex cores is consistent with the appearance of a $d$-wave component in magnetic fields.

In conclusion we have argued that in $A15$ superconductors an attractive $d$-wave pairing channel exists in addition to the $s$-wave pairing channel. This $d$-wave pairing channel accounts for the sharp Raman resonance seen in the $s$-wave superconducting state. Furthermore, this $d$-wave pairing is induced by magnetic fields and can account for the observed anisotropy in the upper critical field and the observation of a sharp reduction of the superconducting coherence length at a hexagonal to square vortex lattice transition that appears at anomalously low fields.

\acknowledgements{This work was supported by NSF grant DMR-0906633. We thank Steve Kivelson, Sonny Rhim, Manfred Sigrist, Kazuo Ueda, and Michael Weinert for useful discussions. }


\begin{references}
\bibitem{davis} J.P. Davis et al. Nature Physics {\bf4}, 571 (2008).

\bibitem{sauls} M. Fogelstr$\ddot{o}$m, D. Rainer, J. A.
    Sauls, Phys. Rev. Lett. {\bf 79}, 281 (1997).

\bibitem{kuroki} K. Kuroki et al. Phys. Rev. Lett. {\bf101}, 087004 (2008).

\bibitem{lee} W. C. Lee, S. C. Zhang, and C. Wu, Phys. Rev. Lett. {\bf102}, 217002
    (2009).

\bibitem{hackl} R. Hackl and R. Kaiser, J. Phys. C: Solid State Phys. {\bf21}, L453 (1988).

\bibitem{math} B. T. Matthias et al. Phys. Rev. {\bf95}, 1435 (1954).

\bibitem{geller} S. A. Geller, Acta Cryst. {\bf9}, 885 (1956).

\bibitem{bat1} B. W. Batterman, Phys. Rev. Lett. {\bf 13}, 390 (1964).

\bibitem{shirane} G. Shirane, Phys. Rev. B {\bf 4}, 2957 (1971).

\bibitem{labb} J. Labbe and J. Friedel, J. Phys. Radium {\bf27}, 153 (1966).

\bibitem{Gorkov} L. P. Gorkov, Sov. Phys. JETP {\bf9}, 1364 (1959).

\bibitem{matd} L. F. Mattheiss, Phys. Rev. B {\bf12}, 2161 (1975).

\bibitem{mihai} T. W. Mihalisin and R. D. Parks, in Low Temperature Physics,
 LT9, edited by J. B. Daunt, D. O. Edwards, F. J. Milford, and M. Yaqub
 (Plenum, New York, 1965).

\bibitem{jarlborg} T. Jarlborg, A. A. Manuel, and M. Peter, Phys. rev. B {\bf27}, 4210 (1983).

\bibitem{leggett} A. J. Leggett, Prog. Theor. Phys. {\bf36}, 901 (1966).

\bibitem{kumar} P. Kumar and P. Wolfle, Phys. Rev. Lett. {\bf59}, 1954 (1987).

\bibitem{matb} L. F. Mattheiss and W. Weber, Phys. Rev. B {\bf25}, 2248 (1982).

\bibitem{bhatt} R. N. Bhatt, Phys. Rev. B {\bf16}, 1915 (1977).

\bibitem{pint} L. Pintschovius, H. Takei, and N. Toyota, Phys. Rev. Lett. {\bf54}, 1260 (1985).


\bibitem{geerk} J. Geerk, U. Kaufmann, W. Bangert, and H. Rietschel, Phys. Rev. B {\bf33}, 1621 (1986).

\bibitem{anderson} P. Morel,  and P. W. Anderson, Phys. Rev. {\bf125}, 1263 (1962).


\bibitem{khlopkin} M. N. Khlopkin, JETP Letters {\bf69}, 26 (1999).

\bibitem{Yethiraj} M. Yethiraj et al. Phys. Rev. Lett. {\bf82}, 5112 (1999).

\bibitem{kadano} R. Kadano et al. Phys. Rev. B  {\bf74}, 024513 (2006).

\bibitem{sonier} J. E. Sonier et al. Phys. Rev. Lett. {\bf93}, 017002 (2004).

\bibitem{kogan}  V. G. Kogan, P.  Miranovi\ifmmode \acute{c}\else \'{c}\fi{}, Lj. Dobrosavljevi\ifmmode \acute{c}\else \'{c}\fi{}-Gruji\ifmmode \acute{c}\else \'{c}\fi{}, W. E. Pickett, and D. K. Christen, Phys. Rev. Lett. {\bf 79}, 741 (1997).

\bibitem{sup} See supplementary material at [   ].

\end{references}
\end{document}


\title{Supplementary Material}
\author{S. Mukherjee}
\affiliation{NBIA, Niels Bohr Institute, Blegdamsvej 17, 2100 Copenhagen $\emptyset$, Denmark}
\author{D. F. Agterberg}
\affiliation{Department of Physics, University of Wisconsin-Milwaukee, Milwaukee, WI 53211}
\maketitle

\section{Tight Binding Model}

The unit cell for V$_3$Si contains six effective Vanadium atoms per unit cell. We use a tight binding model that included the $(3z^2-r^2)$ orbitals in the z-direction, and the corresponding orbitals along the x and y directions given by group rotation operations that satisfy cubic symmetry. Density functional theory shows that these orbitals are the most important near the Fermi surafce \cite{matb}. The Vanadium atoms in a unit cell along the x, y, and z chain directions have position $\ver_m=(\hvy/2\pm \hvx/4,\hvz/2\pm \hvy/4,\hvz/2\pm \hvy/4)$ in units of lattice spacing with respect to the center of the unit cell and correspond to orbitals with symmetries $\nu=(3x^2-r^2,3y^2-r^2,3z^2-r^2)$ respectively. We label these 1 through 6 as shown in Fig.~1. We include nearest (intra-chain) and next-nearest neighbor (inter-chain) hopping matrix elements (see Fig.~1). Letting  $c_{m,i}^{\nu}={1\over \sqrt{N}}\sum_ie^{-i\vk\cdot (\veR_i+\ver_m)} c_{m\vk}^{\nu}$ where $c_{m,i}^{\nu}$ annihilates an electron on the $m^{th}$ Vanadium atom that has the symmetry of the $\nu$ orbital in the unit cell given by the $\veR_i$, the tight binding Hamiltonian in reciprocal space is expressed in matrix form $H=\sum_{\vk}{\Psi_{\vk}}^{\dag}M_k\Psi_{\vk}$ with

\begin{eqnarray}
M_k=\left(
      \begin{array}{cccccc}
        -\mu & \epsilon_{x}(\vk) & \epsilon_{13}(\vk) & \epsilon_{14}(\vk) & \epsilon_{15}(\vk) & \epsilon_{16}(\vk) \\
        \epsilon_{x}(\vk) & -\mu & \epsilon_{23}(\vk) & \epsilon_{24}(\vk) & \epsilon_{25}(\vk) & \epsilon_{26}(\vk) \\
        \epsilon_{13}^{*}(\vk) & \epsilon_{23}^{*}(\vk) & -\mu & \epsilon_{z}(\vk) & \epsilon_{35}(\vk) & \epsilon_{36}(\vk) \\
        \epsilon_{14}^{*}(\vk) & \epsilon_{24}^{*}(\vk) & \epsilon_{z}(\vk) & -\mu & \epsilon_{45}(\vk) & \epsilon_{46}(\vk) \\
        \epsilon_{15}^{*}(\vk) & \epsilon_{25}^{*}(\vk) & \epsilon_{35}^{*}(\vk) & \epsilon_{45}^{*}(\vk) & -\mu & \epsilon_{y}(\vk) \\
        \epsilon_{16}^{*}(\vk) & \epsilon_{26}^{*}(\vk) & \epsilon_{36}^{*}(\vk) & \epsilon_{46}^{*}(\vk) & \epsilon_{y}(\vk) & -\mu \\
      \end{array}
    \right)
\end{eqnarray}
and $\Psi_{\vk}=(c_{1k}^{\gamma},c_{2k}^{\gamma},c_{3k}^{\gamma},c_{4k}^{\gamma},c_{5k}^{\gamma},c_{6k}^{\gamma})$. In the above $6\times6$ matrix $\mu$ is the chemical potential that is adjusted to yield the correct qualitative form for the known Fermi surface sheets as evaluated from the full DFT calculations\cite{jarlborg}. Here the terms $\epsilon_{\alpha}(\vk)=2t_1\cos(k_{\alpha}/2)$ where $\alpha=\{x,y,z\}$ and the interchain terms are given by
\begin{eqnarray}
\epsilon_{24}(\vk)&=&2t_2e^{-i(k_x/4)}e^{-i(k_z/4)}\cos(k_y/2) \nonumber\\
\epsilon_{23}(\vk)&=&2t_2e^{-i(k_x/4)}e^{i(k_z/4)}\cos(k_y/2)\nonumber\\
\epsilon_{14}(\vk)&=&2t_2e^{i(k_x/4)}e^{-i(k_z/4)}\cos(k_y/2)\nonumber\\
\epsilon_{13}(\vk)&=&2t_2e^{i(k_x/4)}e^{i(k_z/4)}\cos(k_y/2)\nonumber\\
\epsilon_{26}(\vk)&=&2t_2e^{-i(k_x/4)}e^{-i(k_y/4)}\cos(k_z/2)\nonumber\\
\epsilon_{25}(\vk)&=&2t_2e^{-i(k_x/4)}e^{i(k_y/4)}\cos(k_z/2)\nonumber\\
\epsilon_{16}(\vk)&=&2t_2e^{i(k_x/4)}e^{-i(k_y/4)}\cos(k_z/2)\nonumber\\
\epsilon_{15}(\vk)&=&2t_2e^{i(k_x/4)}e^{i(k_y/4)}\cos(k_z/2)\nonumber\\
\epsilon_{46}(\vk)&=&2t_2e^{-i(k_z/4)}e^{-i(k_y/4)}\cos(k_x/2)\nonumber\\
\epsilon_{45}(\vk)&=&2t_2e^{-i(k_z/4)}e^{i(k_y/4)}\cos(k_x/2)\nonumber\\
\epsilon_{36}(\vk)&=&2t_2e^{i(k_z/4)}e^{-i(k_y/4)}\cos(k_x/2)\nonumber\\
\epsilon_{35}(\vk)&=&2t_2e^{i(k_z/4)}e^{i(k_y/4)}\cos(k_x/2).
\end{eqnarray}

\section{Derivation of Ginzburg Landau Theory}
In addition to the tight binding Hamiltonian above, a superconducting instability is present in the system through the following quartic interaction term,

\begin{eqnarray}
H_{BCS}=-V_0\sum_{\gamma,m,i}(c_{m,i,\uparrow}^{\gamma})^{\dag}(c_{m,i,\downarrow}^{\gamma})^{\dag}
c_{m,i,\downarrow}^{\gamma}c_{m,i,\uparrow}^{\gamma}
\end{eqnarray}
For simplicity the BCS Hamiltonian has been approximated as an onsite interaction which is a reasonable approximation owing to the intra-chain phonons being the most strongly coupled phonons\cite{pint}.

Fourier transforming and diagonalizing the band part of the Hamiltonian gives a momentum dependance to the quartic interaction.
In terms of the eigenstates of the band Hamiltonian, the full Hamiltonian can be expressed in the momentum space,
\begin{eqnarray}
H&=&\sum_{\alpha=[1\rightarrow6]}\epsilon_\alpha(\vk) a_\alpha^{\dag}(\vk)a_\alpha(\vk) +\sum_{\alpha,\beta} V_{\alpha,\beta}(\vk,\vk')a_\alpha^{\dag}(\vk+\vq/2,\uparrow)a_\alpha^{\dag}(-\vk+\vq/2,\downarrow) a_\beta(-\vk'+\vq/2,\downarrow)a_\beta(\vk'+\vq/2,\uparrow)
 \end{eqnarray}
 where $\alpha,\beta$ represent the eigenstates of the diagonalized band Hamiltonian. If $\hat{c}=\hat{P}^{\dag}\hat{a}$, where $\hat{U}=\hat{P}^{\dag}$ is the transpose of the unitary eigenvector matrix whose components are $U_{ij}$ then the momentum dependance of the quartic interaction term is
 \begin{eqnarray}
 V_{\alpha,\beta}(\vk,\vk')&=&-V_0\sum_{i}U_{i,\alpha}^{*}(\vk) U_{i,\alpha}^{*}(-\vk)U_{i,\beta}(-\vk')U_{i,\beta}(\vk').
 \end{eqnarray}

\begin{figure}
\epsfxsize=3.0 in \center{\epsfbox{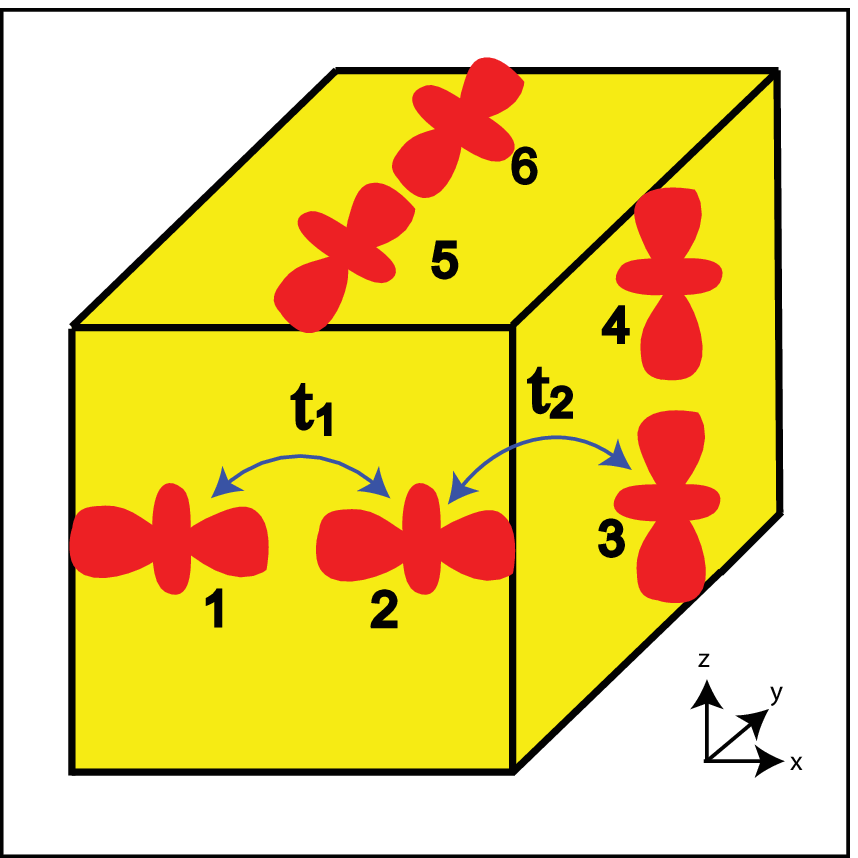}}
\caption{Orbitals centered on Vanadium atoms in
the unit cell and the corresponding hopping matrix elements included in the Hamiltonian.}
\label{fig:eigenvalue}
\end{figure}

The gap function is
$\Delta_{\alpha,\uparrow\downarrow}(\vk,\vq)=\sum_{\vk',\beta}
V_{\alpha,\beta}(\vk,\vk')\langle c_{\vk+\vq/2,\beta,\uparrow}c_{-\vk+\vq/2,\beta,\downarrow}\rangle$
where $\alpha,\beta$ represent band indices and inhomogeneity of the
order parameter is included through the center if mass momentum index $\vq$ (this will allow the GL coefficients $\kappa_i$ to be determined). The superconducting gap equation
becomes
\begin{eqnarray}
\label{eq:gap}
\Delta_\alpha(\vk,\vq)=T\sum_{n,\beta,\vk'}{V_{\alpha,\beta}(\vk,\vk')\Delta_\beta(\vk',\vq)\over
(\tilde{\omega}_{n}^2+\epsilon_\beta^2(\vk')+|\Delta_\beta(\vk')|^2)}
\end{eqnarray}
where $\tilde{\omega}_{n}=(2n+1)\pi T+i\vq\cdot
\nabla_{\vk'}\epsilon_{\beta}(\vk')$.

Within a weak coupling approximation the gap equation close to the superconducting T$_c$ can be integrated to yield the
Ginzburg Landau free energy. The general method for performing this operation is well known\cite{mineev}. Here we assign the order parameters
$\psi_s$ to the $A_{1g}$ representation solution and $(\eta_1,\eta_2)$ to the $E_g$ representation of cubic symmetry.
The superconducting gap can be expressed in terms of these order parameters
in the form,
\begin{eqnarray}
\Delta(\vk)=\psi f_{\psi}(\vk)+\eta_1 f_{\eta_1}(\vk)+\eta_2 f_{\eta_2}(\vk)
\end{eqnarray}
Here $f_{\psi}$ is the $s$-wave basis function and ($f_{\eta_1}$, $f_{\eta_2}$) correspond to the basis functions
for the two dimensional E$_g$ irreducible representation of cubic symmetry. Expanding the
gap equation to second order in $\vq$ and third order in the order parameter yields an equation that can be integrated
in terms of the order parameter to obtain the Ginzburg Landau free energy. Note that the basis functions $f(\vk)$ are
the self consistent numerical solutions of the linear gap equation.

 Therefore using the above procedure and
expanding the superconducting gap equation in the weak coupling limit close to the superconducting transition
 temperature the free energy in the
$s$-$d$ basis is obtained,
 \begin{eqnarray}
&&F=\alpha_s|\psi|^2+\alpha_d|\veta|^2+\kappa_1'|\vD\psi|^2+\kappa_2'(|\vD\eta_1|^2+|\vD\eta_1|^2)+
\kappa_3'[(2(D_z\eta_2)(D_z\eta_2)^{*}-(D_x\eta_2)(D_x\eta_2)^{*}-(D_y\eta_2)(D_y\eta_2)^{*})\nonumber\\
&&-(2(D_z\eta_1)(D_z\eta_1)^{*}-(D_x\eta_1)(D_x\eta_1)^{*}-(D_y\eta_1)(D_y\eta_1)^{*})
+\sqrt{3}((D_x\eta_1)(D_x\eta_2)^{*}-(D_y\eta_1)(D_y\eta_2)^{*}+c.c)]\nonumber\\
&&+\kappa_4'[{1\over \sqrt{3}}(2(D_z\psi)(D_z\eta_2)^{*}-(D_x\psi)(D_x\eta_2)^{*}-(D_y\psi)(D_y\eta_2)^{*})
+((D_x\psi)(D_x\eta_1)^{*}-(D_y\psi)(D_y\eta_1)^{*})+c.c]\nonumber\\
&&+\beta_1'|\psi|^4+\beta_2'|\veta|^4+\beta_3'|\veta\cdot
\veta|^2+\beta_4'|\psi|^2(|\eta_1|^2+|\eta_2|^2)+\beta_5'[{\psi^{*}}^2(\eta_1^2+\eta_2^2)+c.c]\nonumber \\ &&
+\beta_6'[\psi^{*}(\eta_1|\eta_1|^2-2\eta_1|\eta_2|^2-\eta_1^{*}\eta_2^2)+c.c] \label{eqn:sd}
\end{eqnarray}
Note that this is the most general free energy for the given representation allowed by cubic symmetry up to the fourth order
in the expansion parameters.
The weak coupling fourth order coefficients are numerically evaluated from the following Fermi surface averages
\begin{eqnarray}
\beta_1'&=& {c\over 2}\langle f_{\psi}^4 \rangle_{FS}\nonumber\\
\beta_2'&=& c \langle f_{\eta_1}^2f_{\psi}^2 \rangle_{FS}\nonumber\\
\beta_3'&=& {c\over 2} \langle f_{\eta_1}^2f_{\eta_2}^2 \rangle_{FS}\nonumber\\
\beta_4'&=& 2c\langle f_{\eta_1}^2f_{\psi}^2 \rangle_{FS}\nonumber\\
\beta_5'&=& {c\over 2}\langle f_{\eta_1}^2f_{\psi}^2 \rangle_{FS}\nonumber\\
\beta_6'&=& c\langle f_{\eta_1}^3f_{\psi} \rangle_{FS}\nonumber\\
\end{eqnarray}
here the constant $c=7V_0\zeta(3)/(8\pi^2T_c^2)$. Similarly the gradient term coefficients are given by
\begin{eqnarray}
\kappa_1' &=& c'/3\langle f_{\psi}^2\vv_F^2\rangle_{FS}\\
\kappa_2' &=& c'/3\langle f_{\eta_1}^2\vv_F^2\rangle_{FS}\\
\kappa_3' &=& c'/\sqrt{3}\langle f_{\eta_1}f_{\eta_2}v_{Fx}^2\rangle_{FS}\\
\kappa_4' &=& c'\langle f_{\psi}f_{\eta_1}v_{Fx}^2\rangle_{FS}\\
\end{eqnarray}
where $\vv_F$ is the fermi velocity.

The corresponding free energy in the chain basis has been given in the text. The chain basis is
formally given by
$\psi_x=\psi_s/\sqrt{3}-\eta_1/\sqrt{6}+\eta_2/\sqrt{2}$,
$\psi_y=\psi_s/\sqrt{3}-\eta_1/\sqrt{6}-\eta_2/\sqrt{2}$ and
$\psi_z=\psi_s/\sqrt{3}-\eta_1\sqrt{{2\over 3}}$. The free energy in the chain basis follows from Eq.~\ref{eqn:sd} by the following relations: $\beta_1'=\beta_2'=(\beta_1+\beta_3)/3$,
$\beta_3'=\beta_5'/2=\beta_6'/(2\sqrt{2})=(2\beta_1-\beta_3)/12$, $\beta_4'=(4\beta_1+\beta_3)/3$,
$\kappa_1'=\kappa_2'=(3\kappa_1+2\kappa_2)/5$, $\kappa_3'=-\sqrt{3/8}\kappa_4'=-(3/10)(\kappa_1-\kappa_2)$. The variation of the GL coefficient with the interchain chain hopping is shown in Fig.~2. Note that the largest coefficients are $\beta_1$, $\beta_3$, $\kappa_1$, and $\kappa_2$ and it would be reasonable to simplify the theory including only these terms in a GL analysis (as we have done). A similar simplification does not occur in the $s$-$d$ basis.\\

\begin{figure}
\epsfxsize=6.0 in \center{\epsfbox{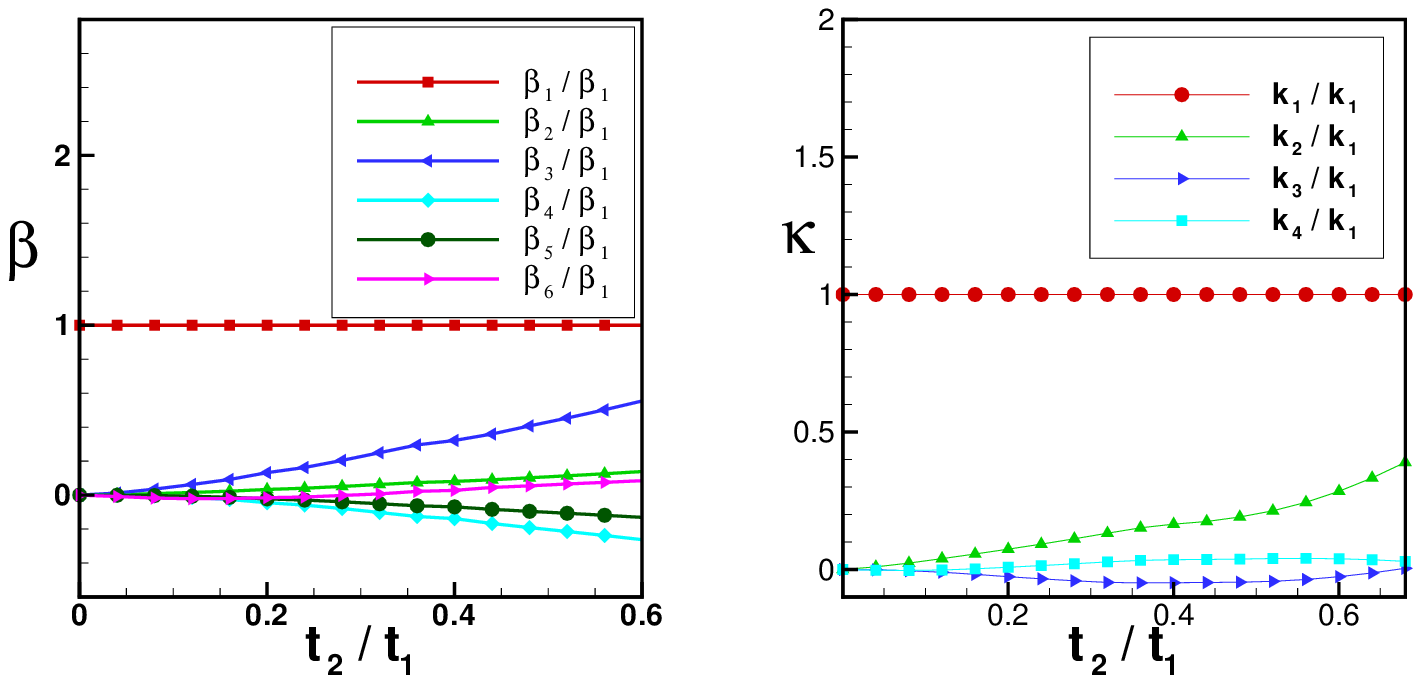}} \caption{Coefficients of the GL theory vs interchain interaction.
The volume enclosed by the Fermi surface has been kept
constant with the variation in $t_2/t_1$.}
\label{fig:gl}
\end{figure}

\section{Upper Critical Field Anisotropy}

The upper critical field may be evaluated by expressing the free energy in terms of the creation and annihilation operators which operate on the Landau levels. For  a magnetic field along the z direction they are defined as, 
 \begin{eqnarray}
 \Pi_{+}={l\over \sqrt{2}}(D_x+iD_y)\\
 \Pi_{-}={l\over \sqrt{2}}(D_x-iD_y)\\
 l^2={2e\over \hbar c}H \nonumber\\
 \end{eqnarray}
 where $l^2={\hbar c\over 2eH}$, with H being an external magnetic field applied along the z direction, and $D_{\alpha}=-i\partial_\alpha-2eA_{\alpha}/c$.
 Expressing the equivalent operators for the different directions of magnetic field we can then write the quadratic portion of the free energy for the magnetic field along the $(1,0,0)$ and $(1,1,0)$ direction in the matrix form as,
 \begin{eqnarray}
 f_{100}=\left(
   \begin{array}{ccc}
     \psi^{*} & \eta_1^{*} & \eta_2^{*} \\
   \end{array}
 \right)
 \left(
                       \begin{array}{ccc}
                      \kappa_1\Pi_{0}+\alpha_s & {\kappa_4\over \sqrt{3}}\Pi_2 & -\kappa_4\Pi_1 \\
                      {\kappa_4\over \sqrt{3}}\Pi_2 & \kappa_2\Pi_{0}-\kappa_3\Pi_2+\alpha_d & -\sqrt{3}\kappa_3\Pi_1 \\
                      -\kappa_4\Pi_1 & -\sqrt{3}\kappa_3\Pi_1 & \kappa_2\Pi_{0}+\kappa_3\Pi_2+\alpha_d \\
                    \end{array}
                  \right)\left(
                           \begin{array}{c}
                             \psi \\
                             \eta_1 \\
                             \eta_2 \\
                           \end{array}
                         \right)\\
                         f_{110}=\left(
   \begin{array}{ccc}
     \psi^{*} & \eta_1^{*} & \eta_2^{*} \\
   \end{array}
 \right)
 \left(
                       \begin{array}{ccc}
                      \kappa_1\Pi_{0} +\alpha_s& {\kappa_4\over \sqrt{3}}\Pi_2 & 0 \\
                      {\kappa_4\over \sqrt{3}}\Pi_2 & \kappa_2\Pi_{0}-\kappa_3\Pi_2+\alpha_d & 0 \\
                      0 & 0 & \kappa_2\Pi_{0}+\kappa_3\Pi_2+\alpha_d \\
                    \end{array}
                  \right)\left(
                           \begin{array}{c}
                             \psi \\
                             \eta_1 \\
                             \eta_2 \\
                           \end{array}
                         \right)\\
 \end{eqnarray}
 Here $\Pi_0=1/l^2(1+2\Pi_{+}\Pi_{-})$, $\Pi_1=1/(2l^2)(\Pi_0+\Pi_{+}^2+\Pi_{-}^2)$, $\Pi_2=1/(2l^2)(\Pi_0-3(\Pi_{+}^2+\Pi_{-}^2))$.
 Note that the (1,1,0) free energy has been written in the rotated basis with,
 \begin{eqnarray}
 x'=z\nonumber\\
 y'=(x-y)/\sqrt{2}\nonumber\\
 z'=(x+y)/\sqrt{2}\nonumber\\
 \end{eqnarray}

 Comparing the upper critical field along the $(1,0,0)$ and $(1,1,0)$ magnetic field directions, we find that
the free energy for the $(1,0,0)$ direction contains additional off diagonal terms which couple the order parameters of the two dimensional representation with $d$-wave symmetry and another term which mixes the $d$-wave and $s$-wave order parameters. Since the presence of these off diagonal coupling terms will enhance the upper critical field in general, the $(1,0,0)$ field direction will have a higher H$_{c2}$ than the H$_{c2}$ along $(1,1,0)$. The role of $s$-$d$ mixing in comparing the H$_{c2}$ for the (1,1,1) and the (1,1,0) magnetic field direction requires a more detailed approach. For $|T-T_c|<<T_s-T_d$, we can use a perturbation theory which treats the Hamiltonian corresponding to the $s$-$d$ mixing as a perturbation (with the $s$-wave state is the ground state). Defining the second order correction ($E^{(2)}$) to the $s$-wave eigenvalue  as
\begin{eqnarray}
{|\alpha_s|\over \kappa_1} = {1\over l^2} + E^{(2)}
\end{eqnarray}
we have
\begin{eqnarray}
E^{(2)}=\sum_m^{'}\left({|<0|H'|\chi_{1m}>|^2\over E_0-E_{1m}}+{|<0|H'|\chi_{2m}>|^2\over E_0-E_{2m}}\right)\\
H'_{110}=k_4/(2l^2)(1+2\Pi_{+}\Pi_{-}+3(\Pi_{+}^2+\Pi_{-}^2))\\
H'_{111}=k_4/l^2(\Pi_{+}^2+\Pi_{-}^2)\\
\end{eqnarray}
In the above expression for the perturbed Hamiltonian, the 2D representation of cubic symmetry has been expressed in the
$\eta_1=(1,\omega,\omega^2)$, $\eta_2=\eta_1^{*}$ basis, where $\omega=e^{2\pi i/3}$. Also the states $|\chi_m>$ are the eigenstates of the unperturbed
Hamiltonian derived from the diagonalization of the $\eta_1$,$\eta_2$ free energy. The solutions can be expressed in terms of
Landau level wave function. For the field along the (1,1,0) direction the eigenstates are

\begin{eqnarray}
\chi_{1m}&=&\phi_m(\epsilon_1x,y/\epsilon_1)\\
\chi_{2m}&=&\phi_m(\epsilon_2x,y/\epsilon_2)\\
\epsilon_1^2&=&\sqrt{{\kappa_1-2\kappa_2\over \kappa_1+\kappa_2}}\\
\epsilon_2^2&=&\sqrt{{\kappa_1+2\kappa_2\over \kappa_1-\kappa_2}}
\end{eqnarray}
where $\phi_m$ represent the $m^{th}$ Landau level solution. Similarly for the field along the (1,1,1) direction the eigenstates $\chi_{1m}$, $\chi_{2m}$ are of the form, $a_m|m>+b_m|m+2>$
where $a_m$ and $b_m$ are known functions. It is found upon solving this problem numerically that the second order correction to the energy will be given by,
\begin{eqnarray}
E^{(2)}_{111}\approx -{8(\kappa_4/\kappa_1)^2\over l^4(T_s-T_d)}\\
E^{(2)}_{110}\approx -{9.5(\kappa_4/\kappa_1)^2\over l^4(T_s-T_d)}
\end{eqnarray}
 therefore the (1,1,0) direction to provides a higher upper critical field than the (1,1,1) magnetic field direction
near T$_c$.
The GL theory given above thus yields the same anisotropy as seen experimentally, {\it independent of the values of the coefficients $\kappa_i$}.